\documentclass[12pt,preprint]{aastex}
\begin{document}

\title{Stellar and Molecular Radii of a Mira Star: First 
Observations with the Keck Interferometer Grism}

\author{J. A.  Eisner,\altaffilmark{1,2} J.R. Graham,\altaffilmark{1}
R.L. Akeson,\altaffilmark{3} E.R. Ligon,\altaffilmark{4} 
M.M. Colavita,\altaffilmark{4} G. Basri,\altaffilmark{1}
K. Summers,\altaffilmark{5} S. Ragland,\altaffilmark{5} \&
A. Booth\altaffilmark{4}}

\altaffiltext{1}{University of California at Berkeley, 
Department of Astronomy, 601 Campbell Hall, Berkeley, CA 94720}
\altaffiltext{2}{Miller Fellow}
\altaffiltext{3}{Michelson Science Center, California Institute of Technology,
MC 100-22, Pasadena, CA 91125}
\altaffiltext{4}{Jet Propulsion Laboratory, California Institute of Technology,
Pasadena, CA 91109}
\altaffiltext{5}{W.M. Keck Observatory, California Association for Research in
Astronomy, 65-1120, Mamalahoa Highway, Kamuela, HI 96743}
\email{jae@astro.berkeley.edu}

\begin{abstract}
Using a new grism at the Keck Interferometer, we obtained spectrally dispersed
($R \sim 230$) interferometric measurements of the Mira 
star R Vir.  These data show that the measured radius of the emission varies
substantially from 2.0-2.4 $\mu$m.  Simple models
can reproduce these wavelength-dependent variations using
extended molecular layers, which absorb stellar radiation and
re-emit it at longer wavelengths.  
Because we observe spectral regions with and without substantial molecular 
opacity, we determine the stellar photospheric radius, uncontaminated
by molecular emission.  
We infer that most of the molecular opacity arises
at approximately twice the radius of the stellar photosphere.  
\end{abstract}


\section{INTRODUCTION \label{sec:intro}}
Miras are asymptotic giant branch (AGB) stars that show 
long-period (hundreds of days) photometric variability 
\citep[e.g.,][]{FEAST+89}.  These stars typically lose mass at a rapid rate
($\sim 10^{-6}$ M$_{\odot}$ yr$^{-1}$), and this
material condenses into heavy elements, molecules, and dust
\citep[e.g.,][]{WILLSON00}.  Since AGB stars account for most of the mass loss 
in the Galaxy, their circumstellar environments are the likely 
factories in which many constituents of planets, and life, are produced.


Separating molecular emission and absorption from stellar photospheric 
emission is important for constraining properties of molecular layers and
the underlying stars.  The opacity and location of 
gaseous layers constrain the temperature structure in the circumstellar
environments of Mira stars; this directly affects the dust condensation radius,
and in turn the mass loss rate.
Measurements of
stellar radii and temperatures uncontaminated by overlying shells
of molecular emission or absorption are needed to estimate the
location of Mira stars in the H-R diagram.  
These quantities also constrain
period-mass-radius relations, with direct
implications for pulsational properties \citep[e.g.,][]{BOWEN88}. 

Historically the circumstellar environments of Mira stars have been
studied through high resolution infrared spectroscopy, which
probed layers of CO and H$_2$O above the stellar photospheres 
\citep[e.g.,][]{HHR82}.  Interferometric imaging, combined with
spectroscopy, has more recently allowed spatially resolved measurements
of radio-wavelength continuum and spectral line emission 
\citep[e.g.,][]{RM97,DK03}, 
mid-IR dust and molecular emission \citep[e.g.,][]{WHT03}, 
and optical/near-IR photospheric and molecular emission 
\citep[e.g.,][]{IRELAND+04,TCV02}. 
These observations showed that molecular emission lies above the stellar
photosphere, and that different molecules lie at different stellocentric radii.

Previous near-IR interferometric observations, which probed the hottest
($\ga 1000$ K) circumstellar emission, used either very low
spectral resolution \citep[$R \la 25$; e.g.,][]{TCV02,MENNESSON+02} 
or a few narrow-band filters designed to probe certain molecular lines 
or nearly line-free regions of the spectrum \citep[e.g.,][]{PERRIN+04b}.  
These observations provided evidence
for different spatial distributions of stellar continuum and molecular
radiation, but lacked the spectral resolution to clearly separate
various components.  With higher spectral resolution ($R \ga 100$), one can
distinguish contributions from H$_2$O, CO, and the stellar photosphere,
and can even discern several of the CO band-heads.
Higher spectral 
resolution data therefore allow an estimate of the relative spatial 
distributions of molecular emission and stellar continuum emission in 
Mira stars. 

Here we present the first observations with a new grism at the Keck
Interferometer, which enables an order of magnitude higher spectral resolution 
($R \sim 230$ across the $K$-band) than previous studies.  
We observed the short-period ($\sim 145$ days) Mira star R Vir (AKA
HD 10994, HR 4808, or IRC 10256).  This object
varies by $\sim 6$ magnitudes in the visible \citep{KHOLOPOV+85} and
$\sim 0.6$ magnitudes in the $K$ band \citep{WMF00}.  The distance to 
R Vir is
$\sim 400$--500 pc \citep{JK92,KNAPP+03}, and the  
mass loss rate is estimated
to be $\sim 10^{-7}$ M$_{\odot}$ yr$^{-1}$ \citep{JURA94}.

\section{EXPERIMENTAL SETUP \label{sec:obs}}
We observed R Vir on UT 2006 May 15 using the Keck Interferometer (KI).
KI is a fringe-tracking long baseline Michelson
interferometer combining light from the two 10-m Keck apertures 
\citep{CW03,COLAVITA+03}.  
We measured $K$-band fringes that were spectrally
dispersed through a replica grism.
The grism has an undeviated wavelength of $2.3$ $\mu$m in the first
order, an apex angle and blaze angle of $6.5^{\circ}$, $21.36$ 
grooves/mm, and an index of refraction of $1.435$ (Infrasil).
The KI detector has a pixel size of 18.5 $\mu$m and a camera with 
a focal length of $\sim 82.5$ mm.  
Thus, the grism provides a spectral resolution of $R=230$ with 42 
10-nm-wide channels across the $K$-band.

We measured squared visibilities ($V^2$) for our target and two calibrator
stars in each of these spectral channels. The calibrator stars are K0-K4
giant stars, with known parallaxes,
whose $K$ magnitudes are within 0.5 mags of the target.  The system
visibility (i.e., the point source response of the interferometer)
was measured with these calibrators, whose angular sizes
were estimated by fitting blackbodies 
to literature photometry.
The calibrator data were weighted by the internal scatter and the temporal 
and angular proximity to the target \citep{BODEN+98}. 
Source and calibrator data were corrected for 
detection biases as described by \citet{COLAVITA99} and averaged into 5-s 
blocks. The calibrated $V^2$ are the averages of 5-s 
blocks in each integration, with uncertainties given by the quadrature addition
of the internal scatter and the uncertainty in the system visibility. 

\begin{figure}[btp]
\plotone{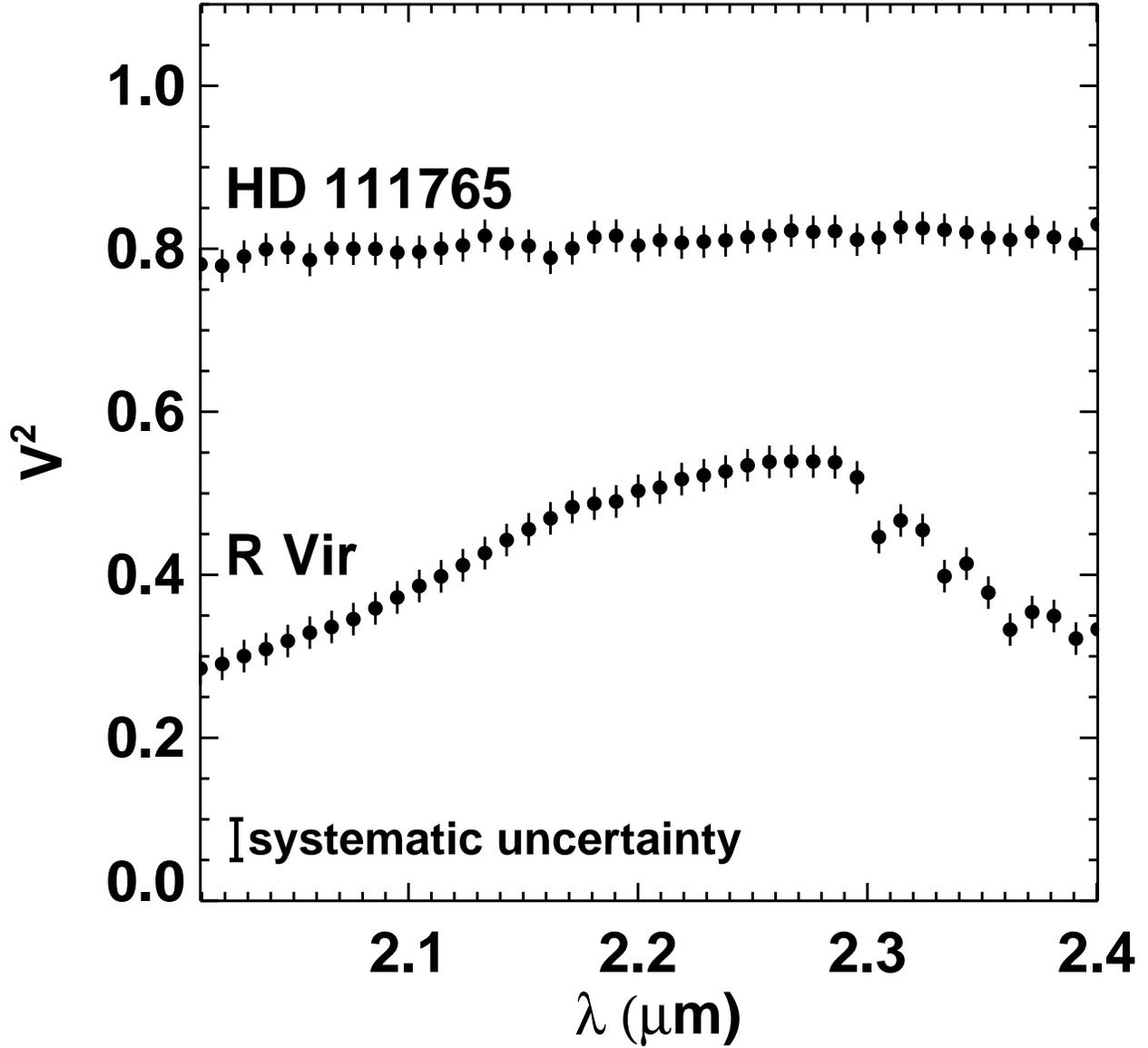}
\caption{Calibrated $V^2$ as a function of wavelength for R Vir and for a 
K4III reference star (note that the reference star is mildly 
resolved and thus $V^2<1$).  The plotted error bars correspond to
channel-to-channel uncertainties of 2\%.  We also indicate a 5\% systematic
uncertainty in the overall calibration that affects the normalization
(\S \ref{sec:obs}).
\label{fig:cals}}
\end{figure}

We averaged our data to produce a single measurement of 
$V^2$ in each spectral channel. Obseravtions of R Vir spanned
0.7 hours, and the averaging has a negligible effect on the
uv coverage.  By comparing our calibrator stars to one another, we see that 
the channel-to-channel variations are small (Figure \ref{fig:cals}).  Using 
HD 107328 (spectral type K0III) to calibrate HD 111765 (spectral type K4III), 
we find a standard deviation of $0.012$ 
in the calibrated $V^2$ versus wavelength.  Calibrating HD 107328 with
respect to HD 111765 we find $\sigma =0.028$.  We therefore adopt 
a channel-to-channel uncertainty of 2\% for our data on R Vir.

The normalization of $V^2$ versus wavelength has larger uncertainty
than the channel-to-channel uncertainties discussed above.  
Observations of a binary star with a known orbit show that 
the calibrated $V^2$ have a systematic uncertainty of $\la 5\%$.
This is a discrepancy in the normalization; if we subtract 0.04 from our
measurements, the measured and predicted $V^2$ are consistent
across all channels at approximately the 1\% level.  
We do not apply this offset to our data, but assume that in addition to the 
2\% channel-to-channel uncertainties described above, the normalization of 
$V^2$ is uncertain by $\sim 5$\%.

We used the counts in each channel observed during 
``foreground integrations'' \citep{COLAVITA99} to recover a crude
spectrum for our target.  We divided the measured flux versus wavelength
for R Vir by the observed flux from an A1V star (smoothed over the Br$\gamma$
feature at 2.16 $\mu$m) and multiplied by a 9000 K 
blackbody to calibrate the spectral
bandpass.  While our spatially filtered data may contain relatively less 
flux at shorter wavelengths due to the larger effect of atmospheric turbulence,
such slopes should be removed through our spectral calibration. 
We estimate that the channel-to-channel uncertainties are 
$\sim 5$--10\%. However there may be larger systematic errors due to 
different $K$ magnitudes of the target star and the A1V spectral calibrator
or changing atmospheric or instrumental conditions between the
source and calibrator observations.  
We normalized our calibrated photometry and then converted to flux units
by assuming a $K$ magnitude of 2.0 based on the current visual phase 
and previously studied $K$-band variability.


\section{MODELING \label{sec:mods}}

\subsection{Star +  Shell \label{sec:mods-simple}}
We begin with a simple model consisting of a star surrounded by a 
single-temperature shell
whose optical depth varies with wavelength (essentially a 
Schuster-Schwarzschild model).  We assume the star emits blackbody
emission from a compact continuum-forming atmosphere.  We ignore limb 
darkening, which may affect the stellar radius 
by $\sim 5\%$ \citep{WELCH94,CDG95}; the effect is comparable 
to the assumed uncertainties in the data.

The free parameters of the model are the stellar angular radius and 
temperature, $\theta_{\ast}$ and $T_{\ast}$,
the shell radius and temperature, $\theta_{\rm shell}$ and $T_{\rm shell}$,
and the optical depth of the shell at each wavelength, $\tau_{\lambda}$.
Since we must determine the best-fit value of $\tau$ at each observed 
wavelength, the model has 46 free parameters.  We measure $V^2$ and flux in 
each spectral channel, and we have 84 data points with which to
constrain these parameters. 

This model has been used previously to model narrow-band interferometric
observations of Mira stars \citep[e.g.,][]{PERRIN+04b,WEINER04}, and can
be used to estimate the relative spatial distributions of
stellar and circumstellar emission.  Moreover, 
comparison of the fitted $\tau_{\lambda}$ with that 
expected for different molecules \citep[based on 
laboratory spectroscopy; e.g.,][]{ROTHMAN+05} constrains which 
molecules are present in the circumstellar environment.

We construct the model by considering sight-lines from the center of the star
out to the limb of the shell, and summing the contributions from each 
line-of-sight.  The specific intensity of the model includes emission from 
both the star and the shell for sight-lines $\theta \le \theta_{\ast}$: 
\begin{equation}
I_{\nu}(\theta) = B_{\nu}(T_{\ast}) {\rm e}^{\left(-\tau_{\lambda}/
\cos \theta\right)} + B_{\nu}(T_{\rm shell}) 
\left[1-{\rm e}^{\left(-\tau_{\lambda}/\cos \theta\right)}\right].
\label{eq:inu1}
\end{equation}
For $\theta_{\ast}<\theta \le \theta_{\rm shell}$, only the shell
contributes:
\begin{equation}
I_{\nu}(\theta) = B_{\nu}(T_{\rm shell}) 
\left[1-{\rm e}^{\left(-2\tau_{\lambda}/\cos \theta\right)}\right].
\label{eq:inu2}
\end{equation}
We compute the flux and normalized visibility for a series of annuli of 
infinitesimal angular width \citep[see e.g.,][]{EISNER+04},
and sum over all annuli to obtain $F_{\nu}$ and $V^2$ for the model.
We fit our data to the modeled $F_{\nu}$ and $V^2$ using a 
Levenberg-Marquardt non-linear least squares fitting algorithm.


\subsection{Star + Molecular Layer \label{sec:model2}}
Instead of considering a shell with arbitrary optical depth 
$\tau_{\lambda}$ (and 42 associated free parameters), we can use molecular 
opacities calculated for stellar atmosphere models (Hauschildt private 
communication).  We replace $\tau_{\lambda}$ with
$\tau_{\rm mol}=N_{\rm col} \times \sigma_{\rm mol}(T_{\rm mol})$, where 
$\sigma_{\rm mol}$ is the molecular cross section (as a function of wavelength)
for a single layer in the Hauschildt atmospheric model,
and $N_{\rm col}$ is the column density of the layer. This column density 
assumes a mix of different molecules (H$_2$, CO, and H$_2$O, among others), 
and the relative abundances are determined through self-consistent modeling of
a cool stellar atmosphere.  Since the relative abundances of H$_2$ and 
H$_2$O and CO may be lower in a shell than in a dense atmosphere, our
fitted column density for the shell may be over-estimated.
The free parameters for this model
are $R_{\ast}$ and $T_{\ast}$,
the molecular layer radius and temperature, $R_{\rm mol}$ and $T_{\rm mol}$,
and the column density of the molecules within the layer, $N_{\rm mol}$.


\section{RESULTS AND CAVEATS}
\begin{figure}[tbhp]
\plottwo{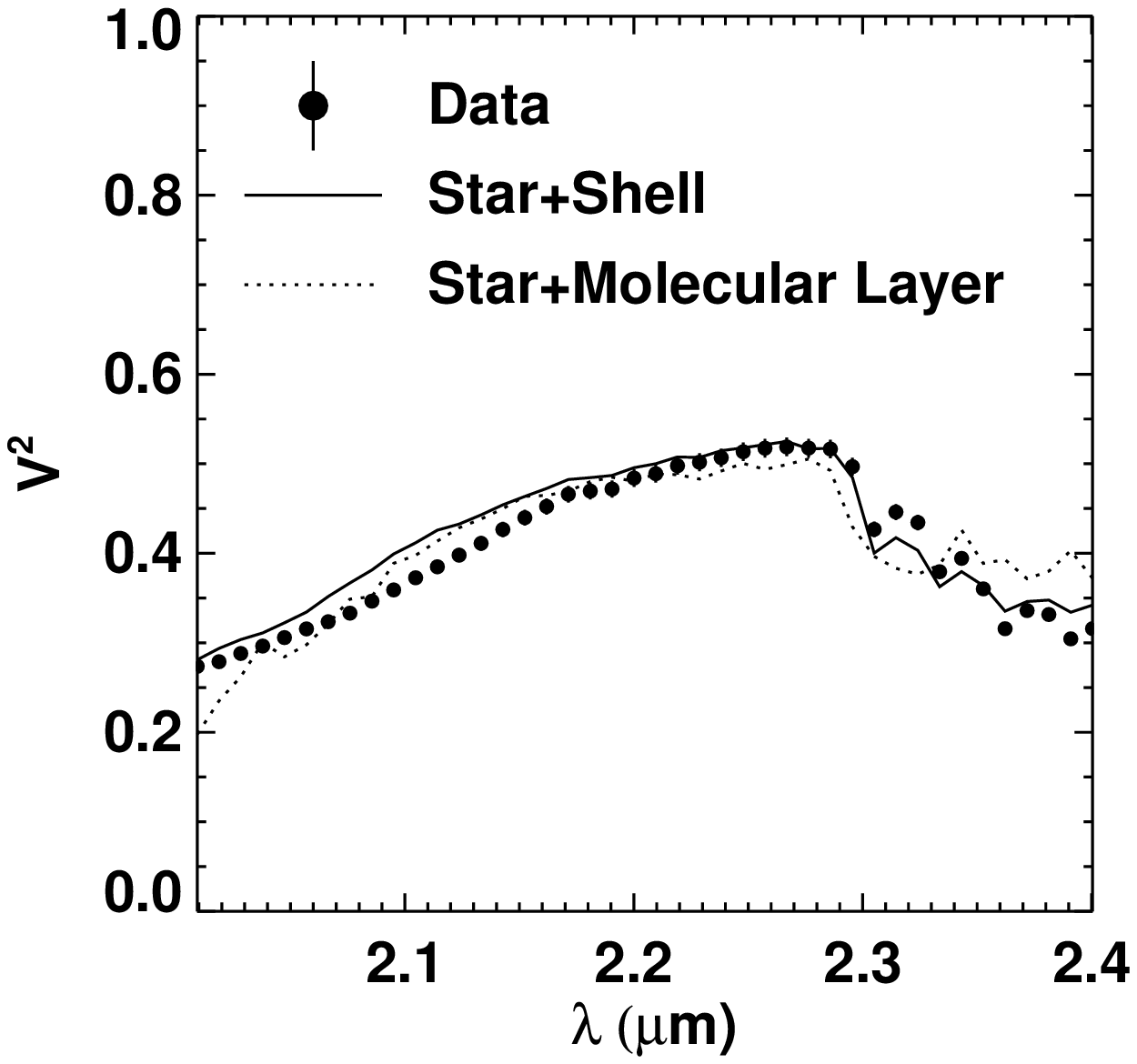}{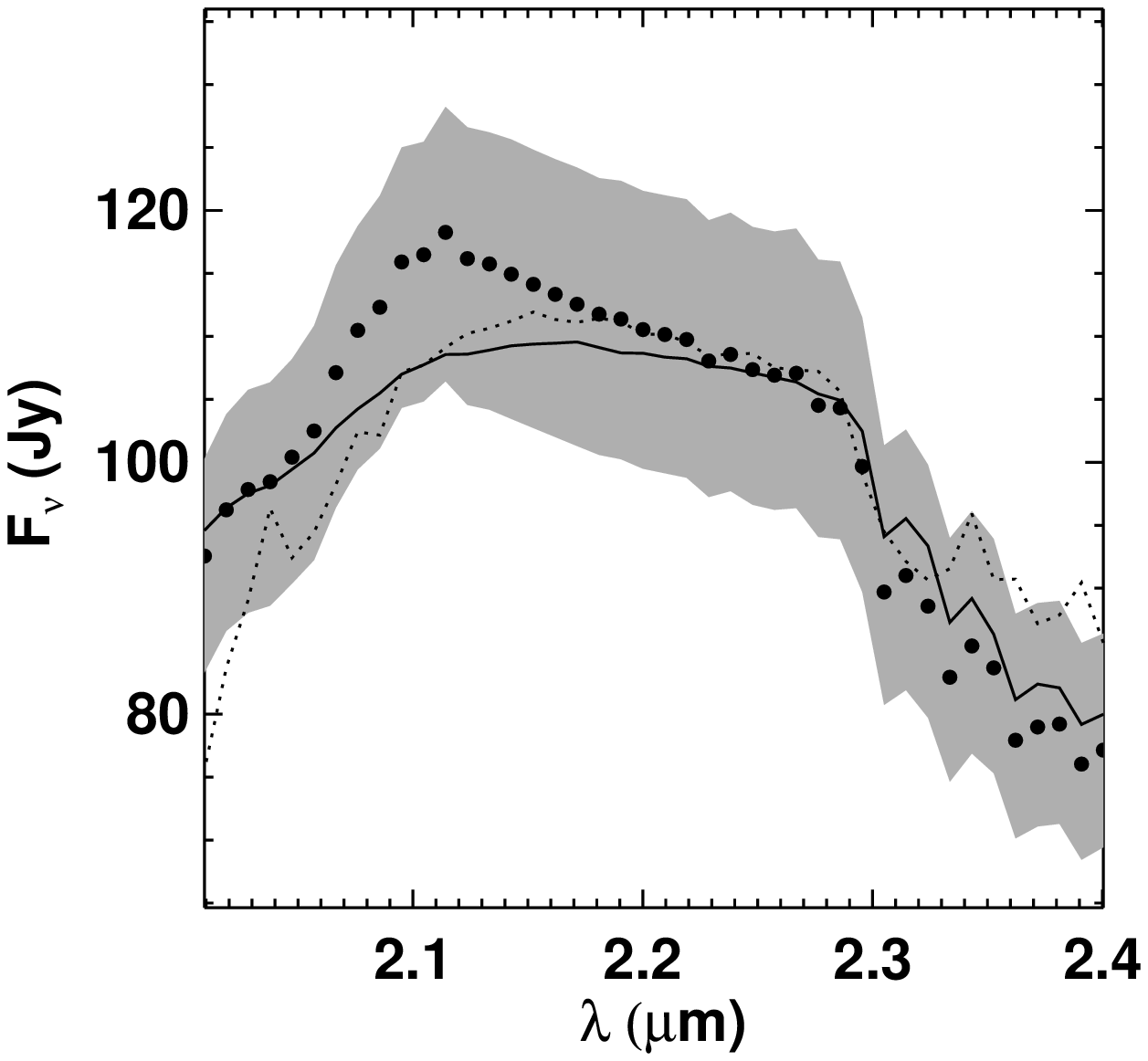}
\caption{Observed $V^2$ and fluxes plotted with the predictions of the
best-fit models.  The shaded region shows assumed 10\% uncertainties for
our flux measurements.  Solid and dashed lines show the predictions for 
the star+shell (\S \ref{sec:mods-simple}) and star+molecular layer
(\S \ref{sec:model2}) models, respectively.
Water vapor lines blend together at the spectral resolution of these 
observations, and contribute substantial opacity from $\sim 2.0$--$2.2$ and 
$\sim 2.3$--$2.4$ $\mu$m \citep[e.g.,][]{LUDWIG71}.  Ro-vibrational lines of
CO contribute opacity between $\sim 2.3$--2.4 $\mu$m 
\citep[e.g.,][]{ROTHMAN+05}, and the four dips in the 
$V^2$ and fluxes between $\sim 2.3$--2.4 $\mu$m occur at the approximate 
wavelengths of the CO overtone bandheads.
\label{fig:shellfits}}
\end{figure}

Our modeling indicates that the stellar radius is $\sim 1.4$ mas
($\sim 130$ R$_{\odot}$ assuming a distance of 450 pc) 
and the stellar temperature is $\sim 3700$--3800 K  
(Table \ref{tab:models}; Figure \ref{fig:shellfits}).  The shell
of molecular material is cooler and located at a larger radius:
$R_{\rm mol} \sim 2.4$ mas 
($\sim 230$ R$_{\odot}$)
and $T_{\rm mol} \sim 1800$--1900 K.  

\begin{figure}[tbhp]
\plotone{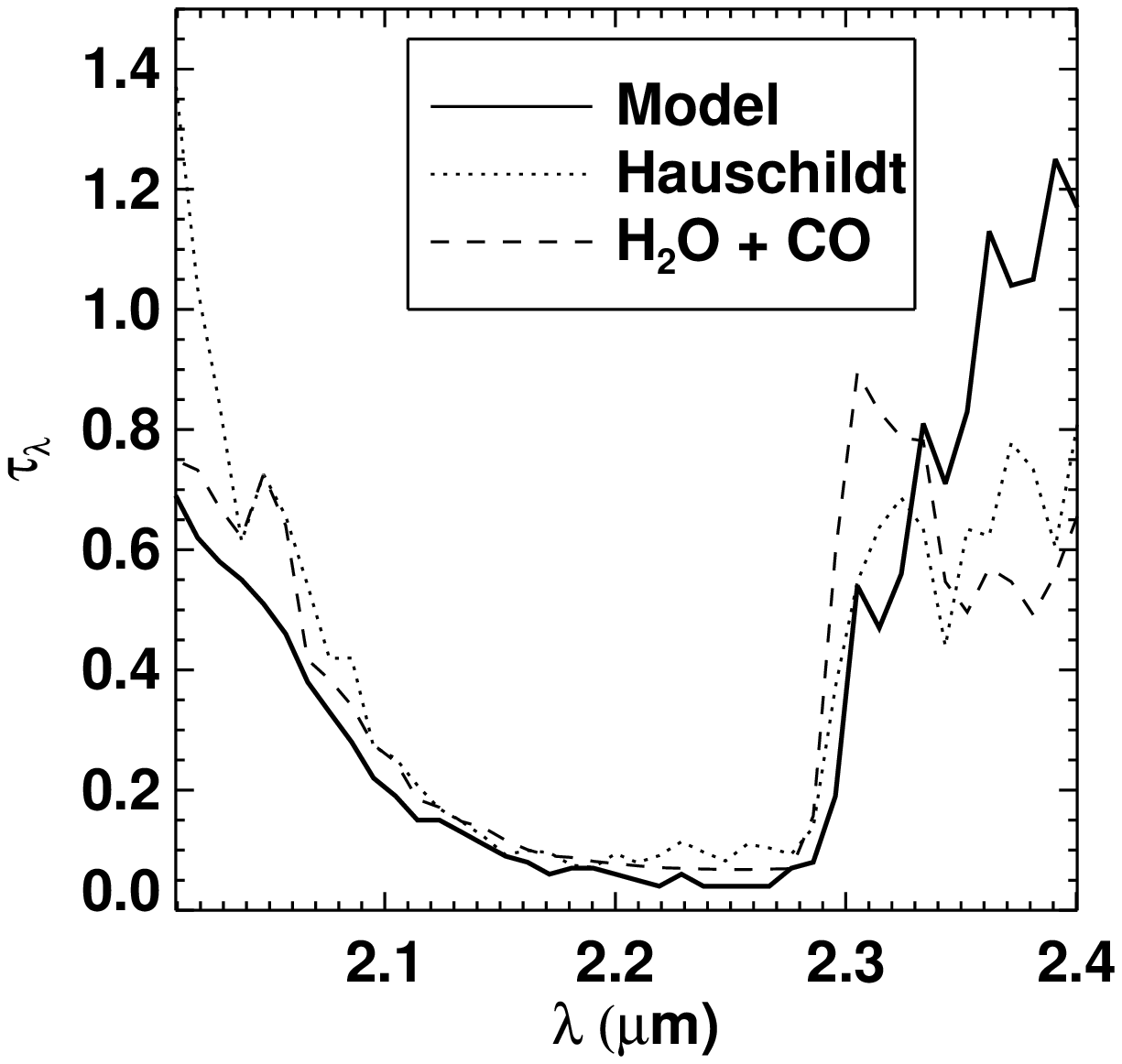}
\caption{Fitted $\tau_{\lambda}$ for the star+shell model
(solid line).  We also plot $\tau_{\lambda}$ computed for a stellar 
atmosphere model  (Hauschildt private communication) in a layer with $T=1800$ K
and $N_{\rm mol}=5$ g cm$^{-2}$, and $\tau_{\lambda}$
based on line-lists of H$_2$O \citep{LUDWIG71} and CO 
\citep[from HITEMP;][]{ROTHMAN+05} 
with $T=1800$ K, $N_{\rm H_2O}=5\times 10^{20}$ cm$^{-2}$, and
$N_{\rm CO}=10^{22}$ cm$^{-2}$ (dashed line).  These column 
densities are not fitted, but rather estimated in an order of magnitude
sense to produce $\tau_{\lambda}$ close to the modeled value.
\label{fig:taus}}
\end{figure}

We have assumed in our models that the molecular material
around R Vir resides in an infinitesimally thin shell.
However material is probably distributed over a
range of radii, and different molecules at different temperatures
may contribute to the optical depth.  Our simple assumption may be 
partially responsible for differences between the modeled opacity and that 
expected from molecular line-lists (Figure \ref{fig:taus}). 
Deviations from spherical symmetry, not included in our models, may 
also contribute to differences between models and data.

The inferred temperature of the molecular shell is substantially
higher than the expected condensation temperature for dust around Mira stars
\citep[$\sim 500$ K; e.g.,][]{RH82,HASHIMOTO94}.
Dust at larger radii, but within the 50 mas ($\sim 23$ AU)
field of view of our observations could contribute some over-resolved flux, 
which would reduce the measured $V^2$ and increase the measured flux.
This should be a minor effect given the small amount of $K$-band flux from
any cool extended dust relative to that in the hot, compact molecular shell
modeled above \citep[see e.g.,][]{EISNER+04}.

\section{DISCUSSION}

Our analysis suggests that the molecules (predominantly H$_2$O)
responsible for most of the opacity in the circumstellar environment of R 
Vir form at approximately twice the radius of the stellar photosphere. The 
molecular material cannot be supported by hydrostatic forces at this radius. 
Propagating shocks associated with the pulsation of Mira stars can lead
to molecular densities at these radii orders of magnitude larger than predicted
by hydrostatic models \citep[e.g.,][]{BOWEN88,WOITKE+99}, and shocks thus 
provide a mechanism for lifting the molecular layer 
in R Vir to its observed location.  

The radius and temperature of the molecular shell imply
radiative equilibrium with the stellar continuum radiation, suggesting that
if the shell lies in a post-shock region, it has had time to cool.
At the molecular densities
in such a shell the radiative cooling time is typically longer than the time 
between successive shocks.  It is more likely that the molecular material in
R Vir cooled via expansion; in a rapidly expanding shell associated with a 
high mass loss rate the cooling time is shorter than the time between
shocks \citep[e.g.,][]{BOWEN88}.


Our model fits imply a stellar radius and temperature of 
$\sim 130$ R$_{\odot}$ and $\sim 3800$ K.
Previous broadband measurements of Mira diameters \citep[e.g.,][]{HST95}, 
which measured contributions from stellar photospheres and molecular layers, 
found large stellar radii, which yielded period-mass-radius relations that 
were consistent with overtone mode pulsation
\citep[e.g.,][]{FEAST96}.  In contrast, the smaller radius 
determined in this work \citep[and radii determined for other Miras from 
spectrally dispersed or narrow band interferometric observations; 
e.g.,][]{PERRIN+04b,WEINER04} is consistent with
the period-mass-radius relationship expected for fundamental mode pulsations, 
compatible with other lines of argument that Miras should pulsate in the
fundamental mode \citep[e.g.,][]{BOWEN88,WOOD+99}.

A logical next step is to compare spectrally dispersed interferometry data to 
dynamic, self-consistent models of Mira atmospheres that include the effects of
H$_2$O opacity.  We attempted such comparisons but currently-available
models \citep[e.g.,][]{ISW04,IS06} have longer periods and are either hotter or
have larger photometric variability amplitudes than our
target. These models predict too much flux and not enough H$_2$O to 
match our data.  More appropriate models will become available in the near 
future (Ireland private communication).  Moreover we will obtain additional 
observations of R Vir, enabling a comparison of models and data over
a range of pulsation phases

\acknowledgements
Data presented herein were obtained at the W.M. Keck Observatory from 
telescope time allocated to NASA 
through the agency's scientific partnership with Caltech and 
the University of California. The Observatory was made possible 
by the generous support of the W.M. Keck Foundation.
We thank the entire KI team for making these observations possible, and
acknowledge the role of MSC software in our data acquisition and analysis.
We are indebted to P. Wizinowich, G. Vasisht, and 
L. Hillenbrand for helping to ensure 
the successful implementation of the grism, and to
P. Hauschildt, J. Carr, and M. Ireland for providing 
models and molecular opacities used in our analysis.

\clearpage

\begin{deluxetable}{lccccc}
\tablewidth{0pt}
\tablecaption{Model Parameter Values
\label{tab:models}}
\tablehead{\colhead{Model} & \colhead{$T_{\ast}$} & \colhead{$\theta_{\ast}$}
& \colhead{$T_{\rm shell}$} & \colhead{$\theta_{\rm shell}$} &
\colhead{$N_{\rm mol}$} \\
 & (K) & (mas) & (K) & (mas) & (g cm$^{-2}$)}
\startdata
Star+Shell & $3800 \pm 440$ & $1.37 \pm 0.07$ & $1800 \pm 100$ & 
$2.44 \pm 0.13$ & - \\
Star+Molecular Layer & $3700 \pm 140$ & $1.40 \pm 0.07 $ & $1900 \pm 100$ & 
$2.42 \pm 0.12$ & $4.3 \pm 0.2$ \\
\enddata
\tablecomments{Uncertainties listed in the table are the quadrature sums of
statistical uncertainties in the fits and assumed 5\% systematic errors.}
\end{deluxetable}

\clearpage
\epsscale{1.0}

\epsscale{1.0}

\epsscale{0.7}


\end{document}